\documentstyle[epsfig]{mn}

\newif\ifAMStwofonts

\ifoldfss
  \ifCUPmtlplainloaded \else
    \NewTextAlphabet{textbfit} {cmbxti10} {}
    \NewTextAlphabet{textbfss} {cmssbx10} {}
    \NewMathAlphabet{mathbfit} {cmbxti10} {} 
    \NewMathAlphabet{mathbfss} {cmssbx10} {} 
  \fi
  \ifAMStwofonts
    \ifCUPmtlplainloaded \else
      \NewSymbolFont{upmath} {eurm10}
      \NewSymbolFont{AMSa} {msam10}
      \NewMathSymbol{\upi}     {0}{upmath}{19}
      \NewMathSymbol{\umu}     {0}{upmath}{16}
      \NewMathSymbol{\upartial}{0}{upmath}{40}
      \NewMathSymbol{\leqslant}{3}{AMSa}{36}
      \NewMathSymbol{\geqslant}{3}{AMSa}{3E}

    \fi
  \fi
\fi 

\ifnfssone
  \newmathalphabet{\mathit}
  \addtoversion{normal}{\mathit}{cmr}{m}{it}
  \addtoversion{bold}{\mathit}{cmr}{bx}{it}
  \newmathalphabet{\mathbfit} 
  \addtoversion{normal}{\mathbfit}{cmr}{bx}{it}
  \addtoversion{bold}{\mathbfit}{cmr}{bx}{it}
  \newmathalphabet{\mathbfss} 
  \addtoversion{normal}{\mathbfss}{cmss}{bx}{n}
  \addtoversion{bold}{\mathbfss}{cmss}{bx}{n}
  \ifAMStwofonts
    \ifCUPmtlplainloaded \else
      \UseAMStwoboldmath
      \makeatletter
      \new@mathgroup\upmath@group
      \define@mathgroup\mv@normal\upmath@group{eur}{m}{n}
      \define@mathgroup\mv@bold\upmath@group{eur}{b}{n}
      \edef\UPM{\hexnumber\upmath@group}
      \new@mathgroup\amsa@group
      \define@mathgroup\mv@normal\amsa@group{msa}{m}{n}
      \define@mathgroup\mv@bold\amsa@group{msa}{m}{n}
      \edef\AMSa{\hexnumber\amsa@group}
      \makeatother
      \mathchardef\upi="0\UPM19
      \mathchardef\umu="0\UPM16
      \mathchardef\upartial="0\UPM40
      \mathchardef\leqslant="3\AMSa36
      \mathchardef\geqslant="3\AMSa3E
    \fi
  \fi
\fi 

\ifnfsstwo
  \DeclareMathAlphabet{\mathbfit}{OT1}{cmr}{bx}{it}
  \SetMathAlphabet\mathbfit{bold}{OT1}{cmr}{bx}{it}
  \DeclareMathAlphabet{\mathbfss}{OT1}{cmss}{bx}{n}
  \SetMathAlphabet\mathbfss{bold}{OT1}{cmss}{bx}{n}
  \ifAMStwofonts
    \ifCUPmtlplainloaded \else
      \DeclareSymbolFont{UPM}{U}{eur}{m}{n}
      \SetSymbolFont{UPM}{bold}{U}{eur}{b}{n}
      \DeclareSymbolFont{AMSa}{U}{msa}{m}{n}
      \DeclareMathSymbol{\upi}{0}{UPM}{"19}
      \DeclareMathSymbol{\umu}{0}{UPM}{"16}
      \DeclareMathSymbol{\upartial}{0}{UPM}{"40}
      \DeclareMathSymbol{\leqslant}{3}{AMSa}{"36}
      \DeclareMathSymbol{\geqslant}{3}{AMSa}{"3E}
    \fi
  \fi
\fi 

\ifCUPmtlplainloaded \else
  \ifAMStwofonts \else 
    \def\upi{\pi}
    \def\umu{\mu}
    \def\upartial{\partial}
  \fi
\fi

\title{Formation Mechanisms for Spirals in Barred Galaxies}
\author{}
\author[E.V.Polyachenko and V.L.Polyachenko]
       {E.V.Polyachenko and V.L.Polyachenko\thanks{Present address:
       		Ostrovitynova St. 32-237, Moscow 117647, Russia
       		(e-mail: evgenii@orc.ru). } \\
       Institute of Astronomy, Moscow 109017, Russia}
\date{}

\begin{document}

\maketitle

\label{firstpage}

\begin{abstract}
We consider a scenario of formation of the spiral structure in
barred galaxies. This scenario includes the new non-resonant mechanism of elongation of spirals,
due to the characteristic behaviour of the gravitational potential beyond the principal
spiral arms.
\end{abstract}

The following scheme is considered.
At first a bar forms in the centre (e.g., as a result of
an instability). The bar induces the spiral resonance
responses off the bar ends. If these primary  (principal) spirals
are strong enough, they initiate the formation of the
nearly circular spirals that elongate the primary spirals.
The process of elongation can be repeated.

To describe the response of the galactic disk, one can assume the model
of the disk with circular
orbits. The linearized hydrodynamical equations give a
relation for perturbations of the surface density $\sigma(r)$ and the
potential $\Phi(r)$ (e.g., see Fridman \& Polyachenko, 1984):
\begin{equation}
\sigma(r) = -\frac1{r}\frac{d}{dr}\left(r\varepsilon\frac{d\Phi}{dr}\right)
+ \frac4{r^2}\varepsilon\Phi
+ \frac4{r\omega_*}\Phi\frac{d}{dr}(\varepsilon\Omega).
\label{eq1}
\end{equation}
Here we assumed that all perturbations $\sim\exp(-i\omega t+2i\varphi)$,
$t$ is time, $\omega = 2\Omega_p + i\gamma$, $\Omega_p$ is the pattern speed,
$\gamma$ is the growth rate, $\varepsilon(r) =
\sigma_0(r)/(\omega_*^2 - \kappa^2)$, $\sigma_0(r)$ is the
unperturbed density, $\kappa^2 = 4\Omega^2 + rd\Omega^2/dr$, $\Omega(r)$ is the local angular
velocity, $\omega_*(r) = \omega - 2\Omega(r)$.

The principal spirals constitute the response (1) to the bar potential
$\Phi(r,\varphi)\sim r^{-n}\cos 2\varphi$ ($n > 0$; the bar is
fixed vertically).  At the outer inner Lindblad resonance
(outer ILR; slow bars), Eq. (1) leads to a trailing spiral with
the maximum length in azimuth equal to $\pi$.
If the bar were within the inner ILR, the spiral structure began with the
leading primary spirals (Pasha \& Polyachenko, 1994).

     For the bar that ends near the corotation $r=r_c$ (CR; fast bars),
Eq. (1) gives the following equation for the resonance spiral
(if $|\sigma'_0/\sigma_0| > |\Omega'/\Omega|$):
\begin{equation}
\varphi = \varphi_c(r)+\pi/2, \quad \tan 2\varphi_c(r) = -\frac{\gamma}{2\Omega_c'
(r - r_c)},
\label{eq3}
\end{equation}
where $\Omega_c'= d\Omega/dr|_{r_c}$. This is a trailing spiral off the bar with the
maximum length equal to $\pi/2$. So the length of the principal
spiral arms can differ slow and fast bars: on average, the
latter should be twice shorter than the former.

Fig. 1a shows the barred galaxy
NGC\,1365, along with the overlaid Fourier harmonic of the light
distribution that corresponds to the predominant two-armed
symmetry. The surface density of the material $\sigma_2$ can be written
as $\sigma_2(r,\varphi) = {\mathrm Re}[\sigma(r) \exp(2i\varphi)]
= A(r)\cos[2\varphi - F(r)]$. Functions $A(r)$ and $F(r)$ are displayed in
Fig. 1b. As is seen, the
spiral arms consist of two clearly different parts: (1) fairly
open {\it principal arms} off the bar; (2) adjacent nearly circular
{\it quarter-turn spirals}, each being terminated by
radially-aligned oscillations made of short alternating leading and trailing
spirals. These two parts differ essentially both in amplitude and
pitch angle.

     The very similar spirals can be observed elsewhere, for example, in the normal
galaxies NGC 3631 and NGC 157 in which the spirals exist not only in
the outer region but also in the central region.

    This phenomenon is in fact quite common. The quarter-turn spirals are
essentially the response of the galactic disk to the gravitational potential
of the principal arms, when the potential reveals the characteristic
behaviour (see dotted lines in Fig. 2). This behaviour can be described as a
transition between spiral and multipole regimes.

\begin{figure}

a)

\centerline{\psfig{figure=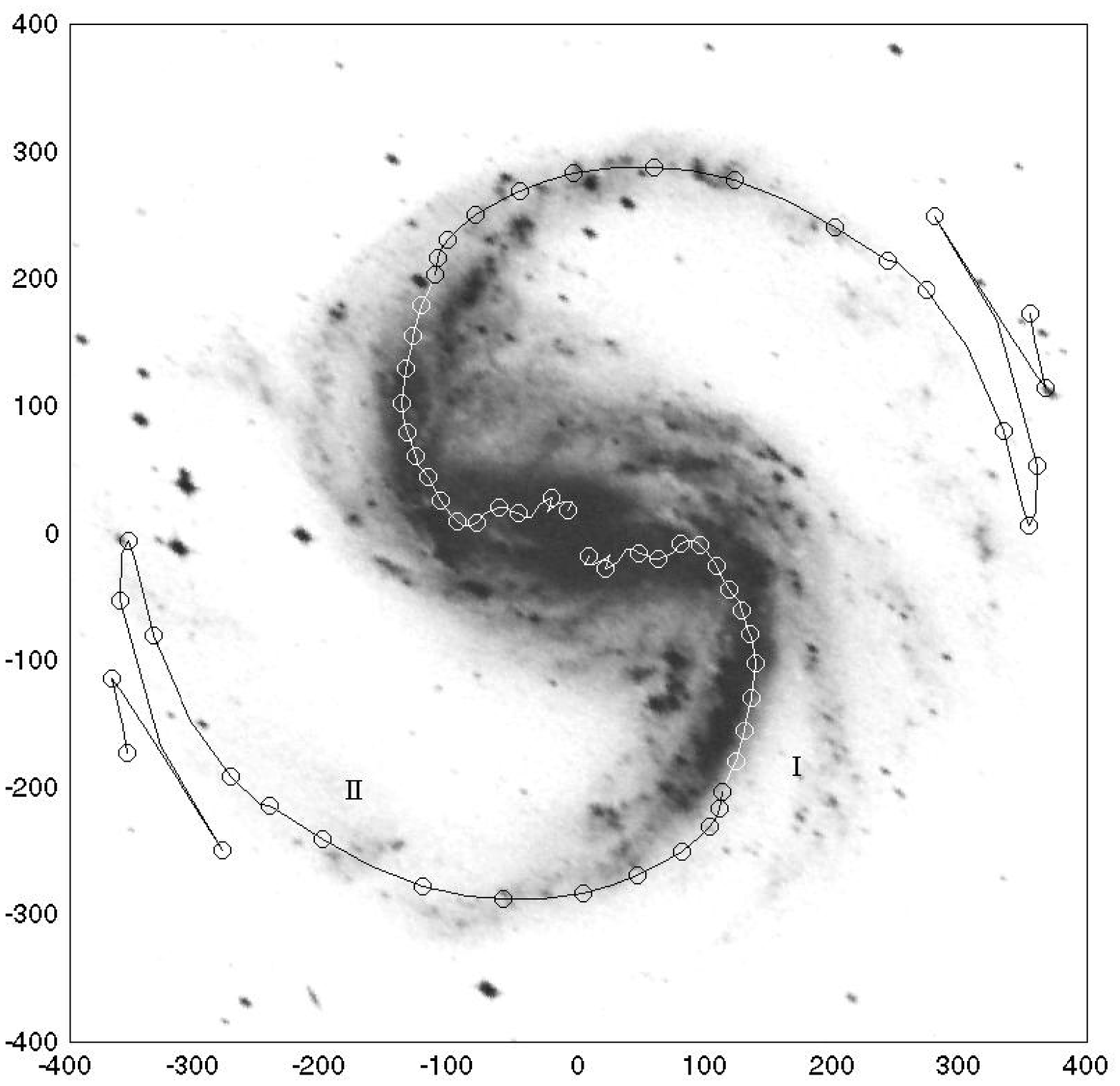,width=7cm}}

b)

\psfig{figure=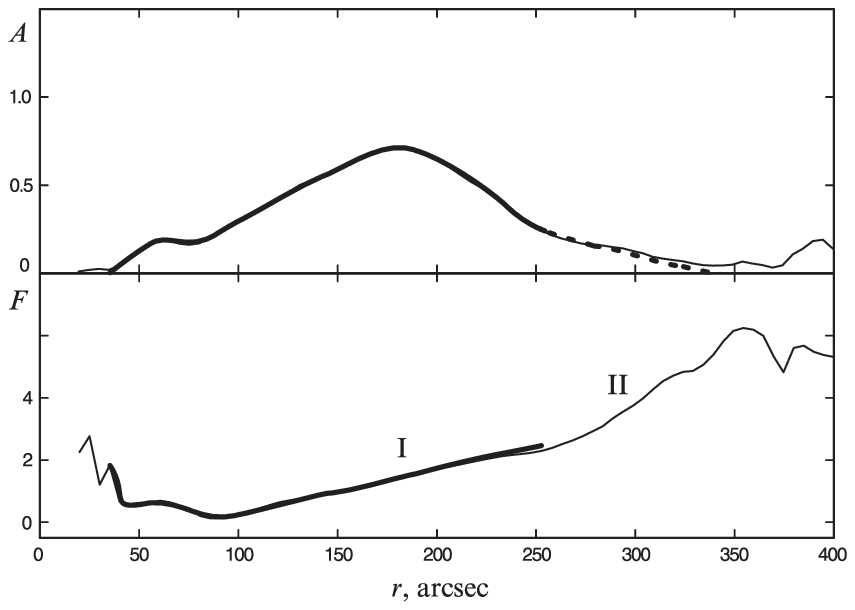, width=8cm}
\caption{ \small
(a) Deprojected image of the galaxy NGC 1365, along with
the overlaid two-armed Fourier harmonic of the light (blue) distribution. The
image is obtained from NED archive. I -- the principal spirals; II -- the
quarter-turn spirals.
(b) Amplitude A(r) and phase F(r) of the Fourier harmonic in Fig.~1a.
Thick lines show the smoothed functions used for calculations of the
galactic disk response (cf. Fig. 2).
}

\label{fig-1}
\end{figure}
\begin{figure}
\centerline{\psfig{figure=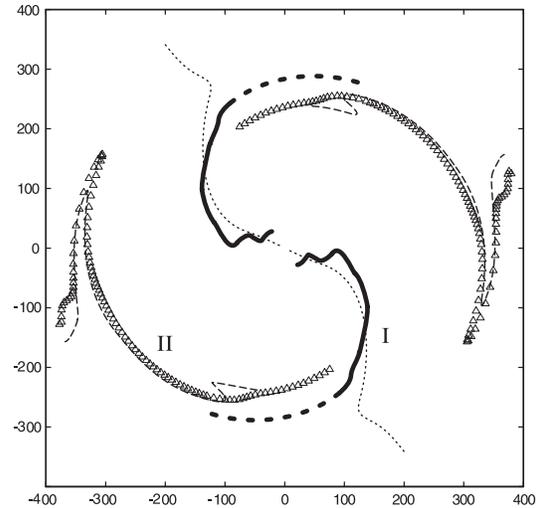,width=7cm}}%
\caption{ \small
Response of the galactic disk of NGC 1365 (triangles, II) to
the gravity of the principal spirals (thick solid curves, I).
As is seen, this response and the quarter-turn spirals in Fig. 1 (a) are very
similar. Dotted lines are used for the curve of minima of the two-armed
potential $\Phi_2(r,\varphi)$. Dashed lines show maxima
of the function ${\mathrm{Re}}[\exp(2i\varphi)\cdot\Phi'']$. Qualitatively, the
patterns of the quarter-turn spirals do not depend on particular values of
disk and spiral wave parameters. In particular, these patterns are
insensitive to a specific band used (although for other bands the amplitude
of the two-armed harmonic can fall in the central region slower).
}
\label{fig-2}
\end{figure}

The potential $\Phi_2(r,\varphi)$ tends to the quadrupole
form well away from the spirals: $\Phi_2(r,\varphi) \rightarrow
r^{-3} \cos 2(\varphi - \varphi_0)$ ($\varphi_0 =$ const).
Hence $\Phi_2(r,\varphi) = {\mathrm Re}[\Phi(r) \exp(2i\varphi)]$ has
non-spiral asymptotic behaviour. In practice, the potential $\Phi_2$
transforms into the multipole form  $\Phi_2(r,\varphi)\sim r^{-n}
\cos 2(\varphi - \varphi_0)$
($n = -d \ln \Phi / d \ln r$, $n \rightarrow 3$ at sufficiently large
radii) well before the quadrupole regime. It occurs in the very
narrow interval of radii just beyond the principal spiral.
Due to fast radial variation in the spiral region as well as in
the transition domain, one can neglect all but one term in (1):
$\sigma(r) \cong -\varepsilon d^2 \Phi/dr^2$. As a rule,
the spirals end within the region, where $\omega_*^2 < \kappa^2$.
Hence $\varepsilon < 0$, so $\sigma(r) \propto\Phi''$.

This result is confirmed by Fig. 2, which shows the loci
of responses, which are calculated by the formula (1), to the
potential produced by the principal arms. The spiral-like potential
leads to $\Phi'' \sim -k^2\Phi$ ($k=F'$) while
$\Phi''\sim +n(n+1)\Phi$ in the multipole regime. It
means that the complex phase of the function $B(r)$, which is the
proportional coefficient between the density and the potential,
$\sigma(r) = B(r)\Phi(r)$, varies from $-\pi$ in the spiral region to
$0$ in the multipole region. Accordingly, the two-armed response has the
angular length equal to $\pi/2$ since the phase change $\Delta F$ is twice as
much as the turn angle $\Delta\varphi$ of the two-armed spiral,
$\Delta F = 2\Delta\varphi$, and
here $\Delta F = \pi$.

\vspace{10mm}
This research has made use of the  NASA/ IPAC Extragalactic Database (NED).
The work was supported in part by grants of RBRF.

\end{document}